%% file: eprint2.tex
\documentclass[12pt]{article}
\usepackage{graphicx}

\newcommand\pubnumber{NuPhys2018-K\"arkk\"ainen\\HIP-2019-12/TH}
\newcommand\pubdate{\today}

\def\napoli{Department of Physics\\
University of Helsinki and Helsinki Institute of Physics,\\ FI-00014 University of Helsinki, FINLAND}

\def\Title#1{\begin{center} {\Large #1 } \end{center}}
\def\Author#1{\begin{center}{ \sc #1} \end{center}}
\def\Address#1{\begin{center}{ \it #1} \end{center}}

\newcommand\pubblock{\rightline{\begin{tabular}{l} \pubnumber\\
         \pubdate  \end{tabular}}}
\newenvironment{Abstract}{\begin{quotation}  }{\end{quotation}}
\newenvironment{Presented}{\begin{quotation} \begin{center} 
             PRESENTED AT\end{center}\bigskip 
      \begin{center}\begin{large}}{\end{large}\end{center} \end{quotation}}
\def\Acknowledgements{\bigskip  \bigskip \begin{center} \begin{large}
             \bf ACKNOWLEDGEMENTS \end{large}\end{center}}

\input econfmacros.tex

\begin{document}
\begin{titlepage}
\pubblock

\vfill
\Title{Neutrino mass via linear seesaw, 331-
model and Froggatt-Nielsen mechanism}
\vfill
\Author{ Katri Huitu, Niko Koivunen and Timo J. K\"arkk\"ainen}
\Address{\napoli}
\vfill
\begin{Abstract}
In this paper, we introduce an  extension of the Standard Model, based on   SU(3)$_\mathrm{C}\times $SU(3)$_\mathrm{L}\times $U(1)$_X$ gauge symmetry (331-model). The 331-models traditionally explain the number of fermion familes in nature. In our model the Froggatt-Nielsen mechanism is incorporated into the 331-setting in a particularly economical fashion.   The model utilizes the  both the Froggatt-Nielsen  and linear seesaw mechanisms to explain the observed fermion mass hierarchies and lightness of neutrinos. In our numerical analysis we found that a $\sim$ 50 TeV new physics scale is able to reproduce correctly all the fermion masses and mixing matrices, including neutrino masses, mass squared differences and mixing matrix.
\end{Abstract}
\vfill
\begin{Presented}
	
	NuPhys2018, Prospects in Neutrino Physics
	
	Cavendish Conference Centre, London, UK,\\ December 19--21, 2018
	
\end{Presented}
\vfill
\end{titlepage}
\def\thefootnote{\fnsymbol{footnote}}
\setcounter{footnote}{0}

\section{Introduction}

The Standard Model (SM) does not  explain the mass and oscillations of neutrinos, number of fermion families or their mass hierarchy. The models based on SU(3)$_\mathrm{C}\times $SU(3)$_\mathrm{L}\times $U(1)$_X$ gauge symmetry (331-models) have been advocated to explain the number of fermion families in nature \cite{Singer:1980sw},\cite{Frampton:1992wt},   and the  Froggatt-Nielsen mechanism (FN) is one of the most popular explanations of the charged fermion mass hierarchy \cite{Froggatt:1978nt}. FN mechanism can be particularly economically incorporated into 331-models (FN331), as was shown in \cite{me}. Here we show that FN331 can be extended to also  explain  the neutrino sector via linear seesaw mechanism \cite{Barr:2003nn}, by introducing three right-handed singlet neutrinos to the model presented in \cite{me}.

\section{Fields and symmetry breaking}
The electric charge in our model is defined as:
\begin{equation}\label{electric charge definition}
Q=T_3-\frac{1}{\sqrt{3}}T_8 + X.
\end{equation}
The left-handed (LH) leptons are assigned into SU(3)$_\mathrm{L}$-triplets and the right-handed (RH) leptons into  SU(3)$_\mathrm{L}$-singlets:
\begin{eqnarray*}
	&&L_{L,i}=\left(\begin{array}{c}
		\nu_{i}\\
		e_{i}\\
		\textsf{$\nu'_{i}$}
	\end{array}
	\right)_{L}\sim \left(\textbf{1},\textbf{3},-\frac{1}{3}\right),\quad 
	  e_{R,i}\sim (\textbf{1},\textbf{1},-1), \quad  N_{R,i}\sim (\textbf{1},\textbf{1},0),
\end{eqnarray*} 
where $i=1,2,3$.
The fields $\nu'_{i}$ and $N_{R,i}$ are new neutrino-like fields.
The quark sector of our model is identical to the one presented in \cite{me}.
 The fermions are taken to be charged under global U(1)$_\mathrm{FN}$-symmetry, which forbids direct inclusion of fermion Yukawa couplings.

Minimal scalar sector that breaks the gauge symmery and produces the tree-level masses to fermions and gauge bosons, consists of three scalar triplets,
\begin{eqnarray*}
	&&\eta=\left(\begin{array}{c}
		\eta^{+}\\
		\eta^{0}\\
		{\eta'}^{+}
	\end{array}
	\right),\quad  
	\rho=\left(\begin{array}{c}
		\rho^{0}\\
		\rho^{-}\\
		{\rho'}^{0}
	\end{array}
	\right),\quad 
	\chi=\left(\begin{array}{c}
		\chi^{0}\\
		\chi^{-}\\
		{\chi'}^{0}
	\end{array}
	\right)\nonumber,
\end{eqnarray*}
with $\eta \sim \left(\textbf{1},\textbf{3},\frac{2}{3}\right)$ and $\rho,\chi \sim \left(\textbf{1},\textbf{3},-\frac{1}{3}\right)$.

Note that  two of the three scalar triplets are in the same representation. This is a special feature of our choice of electric charge in Eq. (\ref{electric charge definition}), and allows us to use the combination $\rho^\dagger \chi$ as the effective flavon. 
The most general electric charge conserving vacuum is,
\begin{equation}
\langle \eta^0\rangle=\frac{v'}{\sqrt{2}},\quad   \langle \rho^0\rangle=\frac{v_1}{\sqrt{2}},\quad  \langle {\rho'}^0\rangle=\frac{v_2}{\sqrt{2}},\quad \langle \chi^0\rangle=\frac{u}{\sqrt{2}}.
\end{equation}
The electroweak symmetry breaking pattern of our model is
$$
\mathrm{SU(3)}_\mathrm{L}\times \mathrm{U(1)}_X\stackrel{u,v_2}{\longrightarrow} \mathrm{SU(2)}_\mathrm{L}\times \mathrm{U(1)}_Y\stackrel{v_1,v'}{\longrightarrow} \mathrm{U(1)}_\mathrm{em},
$$
where $v_2$,$u\sim 50 ~\mathrm{ TeV}$ and  $v_1$,$v'=\mathcal{O}(100~\mathrm{ GeV})$. The new exotic quarks, new scalars  and the new gauge bosons acquire masses  that are of the order of SU(3)$_\mathrm{L}$ breaking scale.

\section{Fusion of FN and seesaw mechanisms}

The combination $\rho^{\dagger}\chi$ is a gauge singlet and  carry a non-zero U(1)$_\mathrm{FN}$ charge. It can therefore play the role of the flavon in the FN mechanism. The Yukawa couplings are forbidden by the U(1)$_\mathrm{FN}$ charge assignment, but the  effective operators 
\begin{equation}
\Delta\mathcal{L}= \sum \limits_{s,f} \left(c_s^{f}\right)_{ij}\left(\frac{\rho^{\dagger}\chi}{\Lambda_{\mathrm{FN}}^2}\right)^{\left(n^s_f\right)_{ij}}\bar{\psi}_{L,i}^f s f_{R,j}+\mathrm{h.c.}
\end{equation}
are allowed. The sum is understood over scalar ($s=\eta,\rho,\chi$) and fermion ($f$) degrees of freedom. $\left(c_s^{f}\right)_{ij}$  are dimensionless and $\mathcal{O}(1)$. The $\bar{\psi}_{L,i}^f$  and $f_{R,j}$ represent here the fermion triplets, anti-triplets and singlets, respectively. Parameter  $\Lambda_\mathrm{FN}$ is the mass scale of the heavy FN messengers that have been integrated out.
Powers $\left(n^s_f\right)_{ij}$ are given by U(1)$_\mathrm{FN}$ charge conservation:
\begin{equation}
\left(n^s_f\right)_{ij}=\left[q\left(\bar{\psi}_{L,i}^f\right)+q\left(f_{R,j}\right)+q(s)\right].
\end{equation}
As the SU(3)$_\mathrm{L}$-symmetry breaks, the "flavon" $\rho^{\dagger}\chi$ acquires vacuum expectation value (VEV), and the Yukawa couplings are generated as effective couplings: 
\begin{equation}
\left(y^f_s\right)_{ij}=\left(c^f_s\right)_{ij}\left(\frac{v_2 u}{2\Lambda^2_\mathrm{FN}}\right)^{\left(n^s_f\right)_{ij}} \equiv \left(c^f_s\right)_{ij}\varepsilon^{\left(n^s_f\right)_{ij}}
\end{equation}
Assuming that $(v_2 u)/\Lambda^2<1$, the fermion mass hierarchy is generated naturally, provided that the U(1)$_{FN}$ charges are adequate. We fix the FN expansion parameter to be the sine of the Cabibbo angle: $\varepsilon \equiv \sin \theta_C \approx 0.23$, since with this choice the CKM-matrix is generated naturally \cite{me}.

Neutrino masses are generated at tree level via linear seesaw mechanism:
\begin{equation}
\mathcal{L}^\nu_\mathrm{mass} = \frac{1}{2}
\left( \begin{array}{ccc}
\overline{\nu}_L & \overline{\nu '}_L & \overline{\left(N_R\right)^c}
\end{array}\right) 
\left( \begin{array}{ccc}
\textbf{0} & 2m^{D\dagger} & m^{N\ast} \\
2m^{D\ast} & \textbf{0} & m^{\prime N\ast}\\
m^{N\dagger} & m^{\prime N \dagger} & M^\ast
\end{array}\right) 
\left( \begin{array}{c}
\nu_L^c \\ \nu_L^{\prime c} \\ N_R
\end{array}\right)+h.c.,
\end{equation}
where all the submatrices have size $3\times 3$, and they are given by,
\begin{eqnarray} 
&&m_{ij}^{\prime N}=\frac{v_2}{\sqrt{2}}c^N_{ij}\epsilon^{q\left(\bar{L}_{L,i}\right)+q\left(N_{R,j}\right)+q(\rho)}
+\frac{u}{\sqrt{2}}{c'}^N_{ij}\epsilon^{q\left(\bar{L}_{L,i}\right)+q\left(N_{R,j}\right)+q(\chi)},
\end{eqnarray}

\begin{equation} 
m^N_{ij}=\frac{v_1}{\sqrt{2}} c^N_{ij}\epsilon^{q\left(\bar{L}_{L,i}\right)+q\left(N_{R,j}\right)+q(\rho)}, \quad 
M_{ij}=\sqrt{\frac{u v_2}{2\epsilon}} c^M_{ij}\epsilon^{q\left(N_{R,i}\right)+q\left(N_{R,j}\right)},
\end{equation}

\begin{equation} 
\textrm{and}\quad m^D_{ij}=\frac{v'}{\sqrt{2}}\left(c^N_{\eta^\ast}\right)_{ij}\epsilon^{q\left(\bar{L}_{L,i}\right)+q\left(\bar{L}_{L,j}\right)-q(\eta)}.
\end{equation}
The light neutrino mass matrix is obtained via block diagonalization:
\begin{equation} 
m_{\nu}=4m^{D\dagger}\left({m'}^{N\dagger}\right)^{-1} M^{\ast} \left({m'}^{N\ast}\right)^{-1} m^{D\ast}
-\left[m^{N\ast}\left({m'}^{N\ast}\right)^{-1}2m^{D\ast}
+2m^{D\dagger}\left({m'}^{N\dagger}\right)^{-1}m^{N\dagger}\right].
\end{equation}

\section{Numerical analysis}
\begin{table}[]
\begin{center} 
	\begin{tabular}{|c|c|c|c|c|c|c|c|c|c|c|c|c|}
		\hline
		\rule{0pt}{3ex}& \multicolumn{3}{c|}{scalars} & \multicolumn{3}{c|}{lepton triplets} & \multicolumn{3}{c|}{lepton singlets} & \multicolumn{3}{c|}{RH neutrinos} \\ \hline
		\rule{0pt}{3ex}Field  & $\eta$  & $\rho$  & $\chi$  & $L_{L,1}$  & $L_{L,2}$  & $L_{L,3}$ & $e_R$ & $\mu_R$ & $\tau_R$ & $N_{R,1}$ & $N_{R,2}$ & $N_{R,3}$ \\ \hline
		\rule{0pt}{3ex}$q_\mathrm{FN}$ & $-1$ & 1 & 0  & \multicolumn{3}{c|}{8} & 3  & 0 & $-2$ & \multicolumn{3}{c|}{0} \\ \hline
	\end{tabular}
\caption{\label{charges}The FN charge assignments for scalar and lepton fields in our model.}
\end{center} 
\end{table}

 Table 1 contains our choice for lepton FN charges. We have used the following benchmark points:
\begin{eqnarray*}
c^L &= \left( \begin{array}{ccc}
1.9753 & -3.6268 & -3.0637 \\ -0.5098 & -3.1787 & -2.0880 \\ -0.93877 & 3.1170 & -0.9170
\end{array}\right), \quad c^N&=\left( \begin{array}{ccc}
3.8705 &-0.5932 &2.5684 \\
0.6689 &-0.8576 &2.6275 \\
 -3.3979 &-4.1881 &4.2437
\end{array}\right),\\
c^M &= \left( \begin{array}{ccc}
	4.0638 & 2.3549 & 1.3775 \\
	2.3549 &-1.1210 &-4.4734  \\
	1.3775 & -4.4734&-2.4107
\end{array}\right), \quad c^{N\prime}&=\left( \begin{array}{ccc}
	2.5372 & 3.2794 & 2.5609\\
	-1.9739 & 2.3864 & 2.6567\\
	-4.0871 &4.7905 &-2.0258
\end{array}\right),
\end{eqnarray*}
$$\textrm{and}\quad \left(c^N_{\eta^\ast}\right) = \left( \begin{array}{ccc}
	0 & 1.4089 & 4.9457\\
	-1.4089 & 0 &  1.5234\\
	-4.9457 & -1.5234 & 0
\end{array}\right).
$$
Note that the $c^M$ is a symmetric and $c^N_{\eta^\ast}$ is an antisymmetric matrix. This benchmark choice produces the correct absolute values of the Pontecorvo-Maki-Nakaga-Sakata (PMNS) matrix within the current $3\sigma$ bounds \cite{Esteban:2018azc}.
$$
\left|U_\mathrm{PMNS}\right|_{ij}=\left( \begin{array}{ccc}
0.8309    &0.5355    &0.1510\\
0.4744    &0.5402    &0.6951\\
0.2907    &0.6491    &0.7029
\end{array}
\right)
$$
It also provides light neutrino masses and mass squared differences consistent with cosmology and as well as atmospheric and solar neutrino experiments. Deviations of standard oscillation formulas are manifested by nonunitarities induced by active-sterile neutrino mixing and also by nonstandard interactions (NSI). The NSIs are mediated by gauge bosons and scalars. Neutral scalars will not contribute to NSI. In both cases the effects are suppressed by the SU(3)$_\mathrm{L}\times$ U(1)$_X$-breaking scale to such a degree that they are unobservable in any conceivable near-future experimental setup.

\Acknowledgements
K. H. acknowledges the H2020-MSCA-RICE-2014 grant no. 645722 (NonMinimalHiggs). T. K. expresses his his gratitude to the organizers of NuPhys2018 conference for their hospitality.

\end{document}

%% file: econfmacros.tex
\def\beq{\begin{equation}}
\def\eeq#1{\label{#1}\end{equation}}
\def\eeqn{\end{equation}}

\def\beqa{\begin{eqnarray}}
\def\eeqa#1{\label{#1}\end{eqnarray}}
\def\eeqan{\end{eqnarray}}

\let\bar=\overbar

\def\Dslash{\not{\hbox{\kern-4pt $D$}}}
\def\dslash{\not{\hbox{\kern-2pt $\del$}}}

\def\msb{{\bar{\ssstyle M \kern -1pt S}}}

